\documentclass[review]{elsarticle}

\usepackage{lineno,hyperref}

\journal{Journal of \LaTeX\ Templates}

\usepackage[utf8]{inputenc} 
\usepackage[T1]{fontenc}
\usepackage{url}
\usepackage{ifthen}
\usepackage{amssymb}
\usepackage[cmex10]{amsmath} 
\usepackage{epsfig}
\usepackage{amsfonts}
\usepackage{pict2e}
\usepackage{color}
\usepackage{tikz}
\usepackage{pgfplots}

\def\qed{\hfill $\blacksquare$}
\newtheorem{Property}{Property}
\newtheorem{Theorem}{Theorem}
\newtheorem{Definition}{Definition}
\newtheorem{Lemma}{Lemma}
\newtheorem{Corollary}{Corollary}

\newtheorem{Remark}{Remark}

\def\C{{\mathcal C}}

\def\Q{{\mathcal Q}}

\def\NN{{\mathbb N}}

\begin{document}

\vspace{8ex}
\begin{frontmatter}

\title{Bounds for the capacity error function for
unidirectional channels with noiseless feedback}

\author{Christian Deppe}
\address{Institute for Communications Engineering\\ 
	                  Technical University of Munich\\
                    D-80333 Munich, Germany\\
                    Email: christian.deppe@tum.de}

\author{Vladimir Lebedev}
\address{Kharkevich Institute for Information Transmission Problems\\
	                  Russian Academy of Sciences\\
										127051 Moscow, Russia\\
										Email: lebedev37@mail.ru}
										
\author{Georg Maringer}
\address{Institute for Communications Engineering\\ 
	                  Technical University of Munich\\
                    D-80333 Munich, Germany\\
                    Email: georg.maringer@tum.de}										
																				


\begin{abstract}
In digital systems such as fiber optical communications, 
	the ratio between probability of errors of type $1\to 0$ and $0 \to 1$ can be large. 
	Practically, one can assume that only one type of error can occur. These errors are
	called asymmetric. Unidirectional errors differ from asymmetric type of errors; here 
	both $1 \to 0$ and $0 \to 1$ type of errors are possible, but in any submitted
	codeword all the errors are of the same type. This can be generalized for the $q$-ary case.
	
	We consider $q$-ary unidirectional channels with feedback and give bounds for the
	capacity error function. It turns out that the bounds depend on the parity of the
	alphabet $q$. Furthermore, we show that for feedback, the capacity error function for
	the binary asymmetric channel is different from the symmetric channel. This is in
	contrast to the behavior of the function without feedback.
	\end{abstract}

\begin{keyword}
error-correcting codes, uni-directional errors, feedback
\end{keyword}

\end{frontmatter}

\section{Introduction}\label{introduction}

Shannon analyzed the capacity of a channel based on probabilistic notions. The arguably easiest example is a binary symmetric channel (BSC) with crossover probability $\varepsilon$. In a probabilistic model the number of errors in a block of symbols can be (having different probability) any number between $0$ and the blocklength $n$. In this paper we analyze the situation when the fraction of erroneous symbols is upper bounded by $\tau = \frac{t}{n}$, where $t$ denotes the maximal number of errors per block. Our focus in this work lies in the analysis of the capacity error function of unidirectional channels.

The analysis in this paper is purely combinatorial. That means that we are not distinguishing between likely and unlikely errors. We only consider whether a certain output may occur given that a certain symbol was chosen at the input of the channel.

The difference between symmetrical, asymmetrical and unidirectional errors is described in~\cite{Weber1989}. A unidirectional channel is composed of two special asymmetric channels. During the transmission of each codeword, one of the two channels is randomly selected. Within the transmission of those codeword symbols, the channel behaves like the selected channel and remains unchanged. After each transmitted codeword, the channel is reselected. Precise definitions are given in Section~\ref{sec:feedback_q_ary}.

Asymmetrical channels model photon communication systems better than symmetrical channels. Due to losses within the transmission medium, photons may not reach the receiver. However, the channel cannot generate photons on its own. Therefore the channel is of asymmetric nature~\cite[1.3]{Weber1989}.
For a transmission without feedback, the binary Z-channel (Figure \ref{fig:binary_Z_channel}) has the same capacity error function as the symmetric channel (Figure \ref{fig:symmetric_channel}). This has already been shown by Bassalygo in~\cite{Ba65}. Asymmetric and unidirectional channels without feedback are analyzed in numerous publications. The major results and references can be found in~\cite{macwilliams1977theory} and~\cite{Weber1989}.
In this paper we give lower and upper bounds on the capacity error function of asymmetric and unidirectional channels in the presence of noiseless feedback. For a wide range of $\tau$, we show that the capacity error function for the unidirectional channel composed out of a Z-channel and an inverse Z-channel is larger compared to the one of the symmetric channel .

\section{Q-ary codes with feedback}\label{sec:feedback_q_ary}
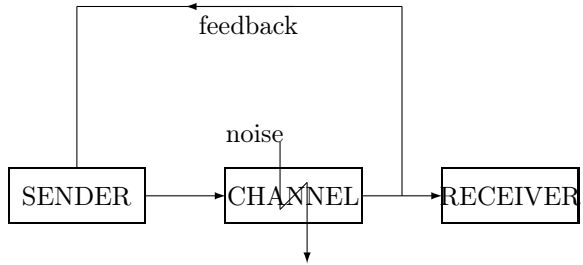
\begin{figure}[h]
	\centering
	\setlength{\unitlength}{0.72 cm}
	\begin{picture}(12,6)
	\put(1,1){\framebox(2.5,1){SENDER}}
	\put(5,1){\framebox(2.5,1){CHANNEL}}
	\put(9,1){\framebox(2.5,1){RECEIVER}}
	\put(3.5,1.5){\vector(1,0){1.5}}
	\put(7.5,1.5){\vector(1,0){1.5}}
	\put(8.25,1.5){\line(0,1){3.5}}
	\put(8.25,5){\vector(-1,0){4}}
	\put(2.25,5){\line(1,0){4}}
	\put(2.25,2){\line(0,1){3}}
	\put(4.5,4.5){feedback}
	\put(5,2.5){noise}
	\put(6,2.5){\line(0,-1){1.25}}
	\put(6,1.25){\line(1,1){0.5}}
	\put(6.5,1.75){\vector(0,-1){1.5}}
	\end{picture}
	\caption{Channel with feedback}
	\label{fig:channel_feedback}
\end{figure}

In our transmission scheme, a sender wants to transmit a finite number of messages
over a channel with noiseless feedback (see Figure \ref{fig:channel_feedback}). 
The discrete channel has a $q$-ary alphabet $\mathcal{X} = \mathcal{Q} = \{0,\dots,q-1\}$ at its input and at its output, $\mathcal{Y} = \mathcal{Q}$. Each message $m$ is encoded into a block of length $n$. For channels without feedback, the codeword $c$ to be transmitted over the channel only depends on the message $m$. In a feedback channel the situation is different.

\begin{Definition}\label{def:feedback_encoding}
\begin{enumerate}
	\item Let the set of possible messages be denoted as $\mathcal{M} = \{1,\dots,M\}$. Then an encoding algorithm for a feedback channel with blocklength $n$ is a set of functions
	
	\begin{equation}
	c_i: \mathcal{M} \times \mathcal{Q}^{i-1} \rightarrow \mathcal{Q}, \quad i \in \{1,\dots,n\} \enspace .
	\end{equation}
	
	The respective coding algorithm is then constructed as
	\begin{equation}
	c(m,y^{n-1}) = ((c_1(m), c_2(m,y_1), \dots, c_n(m,y^{n-1})) \enspace ,
	\end{equation}
	where $y^k = (y_1,\dots,y_k)$.
	\item An encoding algorithm is $(n,M,t)_f$-successful if the corresponding decoder decodes correctly for any transmitted message $m$ and any error pattern with less than or equal $t$ errors caused by the channel.
	\end{enumerate}
\end{Definition}

Definition \ref{def:feedback_encoding} shows that the encoder may adjust its algorithm for sending the $k$th symbol $c_k$ according to the previously received symbols $y^{k-1}$. This extra flexibility can be used to increase the achievable rate of the system.

\begin{Definition}
	Let $2\leq q\in\NN$. The capacity error function $C_q^f(\Gamma,\tau)$ of a channel $\Gamma$ with noiseless feedback denotes the supremum on the rates for which a successful algorithm exists depending on $\tau=\frac{t}{n}$ as the blocklength $n$ goes to infinity. $t$ denotes the maximum number of errors inflicted by the channel noise and $q$ specifies the alphabet size of the channel.
	
	The capacity error function for $\tau=1$ is denoted as $C_{q,0}^f(\Gamma)$ and was introduced by Shannon in~\cite{S56} under the term zero error capacity of the channel.
\end{Definition}
If the sender and the receiver share a channel with noiseless passive feedback (Figure \ref{fig:channel_feedback}), the capacity error function of the symmetric channel is only completely known for the binary case.

\begin{figure}
	\centering
	\begin{tikzpicture}
	\draw (0,0) node [left] {$0$} -- (3,0) node [right] {$0$};
	\draw (0,-2) -- (3,0);
	\draw (0,0) -- (3,-2);
	\draw (0,-2) node [left] {$1$} -- (3,-2) node [right] {$1$};
	\end{tikzpicture}
	\caption{Symmetric Channel $\Gamma^2$}
	\label{fig:symmetric_channel}
\end{figure}
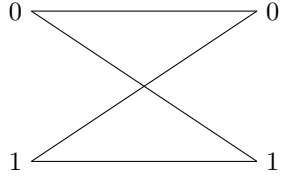

The most frequently analyzed communication channels are of symmetric nature. As an example, we give the previously mentioned symmetric channel $\Gamma^2$ (Figure \ref{fig:symmetric_channel}). In this case it is possible to receive a $1$ when a $0$ was transmitted and a $0$ when a $1$ was transmitted. Both events denote an error, because the correct reception of a symbol is considered to be the event that the sent symbol matches the received symbol. In the binary case, the capacity error function of the symmetric channel was found by Berlekamp~\cite{B68} and Zigangirov~\cite{Z76}
\begin{equation}\label{eq:symmetric_channel}
    C_2^f(\Gamma^2,\tau) = \begin{cases}
	1- h(\tau )& 
	\text{if } 0\leq \tau \leq (3+\sqrt{5})^{-1}\\
	(1-3\tau)\log \left(\frac{1+\sqrt{5}}{2} \right)& \text{if } (3+\sqrt{5})^{-1} \leq \tau \leq \frac{1}{3}\\
	0 & \text{if } \tau > \frac{1}{3}\, .
	\end{cases}
\end{equation}

In general, discrete channels can be specified by bipartite graphs.

\begin{Definition}[Discrete channel/bipartite graph]
	A discrete channel corresponds to a bipartite graph in the following way. Let $\mathcal{V}_{in}$ denote the set of possible input symbols and let $\mathcal{V}_{out}$ denote the set of possible output symbols. Then if it is possible that the channel maps input symbol $i \in \mathcal{V}_{in}$ to output symbol $j \in \mathcal{V}_{out}$, the pair $(i,j)$ is part of the set of edges $\mathcal{E} \subset \mathcal{V}_{in} \times \mathcal{V}_{out}$. Conversely, a bipartite graph between input symbols $\mathcal{V}_{in}$ and $\mathcal{V}_{out}$ defines a discrete channel.
		We denote by $(x_1,x_2,\dots,x_n)$ the sequence of input symbols and by $(y_1,y_2,\dots, y_n)$ the sequence of output symbols, where $x_i$ depends
	on the message and $(y_1,\dots,y^{i-1})$ (see Definition~\ref{def:feedback_encoding}). 
	
	We define $(e_1,e_2,\dots, e_n)$ as the corresponding error vector, where $e_i=y_i-x_i$.
\end{Definition}

Discrete channels with finite input and output alphabets correspond to their respective bipartite graphs by a one to one mapping. Therefore, in this paper we frequently speak about the graphs when we mean the respective channels and vice versa.

\begin{Definition}[Asymmetric channel]
	An asymmetric channel is a discrete channel whose specifying bipartite graph is not the full graph in the sense that the set of edges $\mathcal{E} \neq \mathcal{V}_{in} \times \mathcal{V}_{out}$.
\end{Definition}

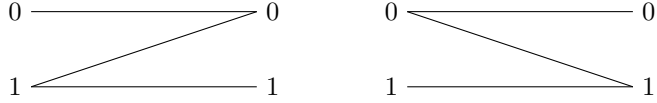
\begin{figure}
	\centering
	\begin{tikzpicture}
	\draw (0,0) node [left] {$0$} -- (3,0) node [right] {$0$};
	\draw (0,-1) -- (3,0);
	\draw (0,-1) node [left] {$1$} -- (3,-1) node [right] {$1$};
	\draw (5,0) node [left] {$0$} -- (8,0) node [right] {$0$};
	\draw (5,0) -- (8,-1);
	\draw (5,-1) node [left] {$1$} -- (8,-1) node [right] {$1$};
	\end{tikzpicture}
	\caption{binary Z-channel and inverse Z-channel}
	\label{fig:binary_Z_channel}
\end{figure}

As an example for an asymmetric channel we propose the Z-channel which is specified by the bipartite graph on the left hand side of Figure \ref{fig:binary_Z_channel}.


\begin{Definition}[Unidirectional channel]
	A unidirectional channel is specified by two asymmetric channels having the same set of input symbols $\mathcal{V}_{in}$ and output symbols $\mathcal{V}_{out}$. One of the channels  allows only positive error vectors $(e_i \geq 0 \; \forall i)$, whereas the other one only allows negative error vectors $(e_i \leq 0 \; \forall i)$. Within each transmission of a codeword, the channel is specified by one of the asymmetric channel. Sender and receiver know both channels, but they do not know to which asymmetric channel the unidirectional channel corresponds. This may change for each codeword.
\end{Definition}

In Figure \ref{fig:binary_Z_channel} the binary unidirectional channel is shown. It is denoted as $\Gamma_U^2$ and is composed of the binary Z-channel and its counterpart, the inverse Z-channel.


\begin{Definition}[Generalized Z/inverse Z-channel]
	The generalized Z-channel $\Gamma_Z^q$ is specified by the bipartite graph with input nodes $\mathcal{V}_{in} = \mathcal{V}_{out} = \{0,\dots, q-1\}$ and the set of edges $\mathcal{E}^q_Z = \{(i,i-1) : i \in \{1,\dots,q-1\}\} \cup \{(i,i) : i \in \{0,\dots,q-1  \}\}$.
	
	The inverse Z-channel $\Gamma_{\reflectbox{$\mathsf{Z}$}}^q$ is specified by the bipartite graph with input nodes $\mathcal{V}_{in} = \mathcal{V}_{out} = \{0,\dots, q-1\}$ and the set of edges $\mathcal{E}^q_{\reflectbox{$\mathsf{Z}$}} = \{(i,i+1) : i \in \{0,\dots,q-2\} \cup \{(i,i) : i \in \{0,\dots,q-1 \}  \}$.
\end{Definition}

\begin{Remark}\label{rem:Z_inverse_Z}
	The capacity error function of the generalized Z-channel $\Gamma_Z^q$ is equal to the capacity error function of the inverse generalized Z-channel $\Gamma_{\reflectbox{$\mathsf{Z}$}}^q$.
\end{Remark}
{\bf Proof:} The inverse Z-channel $\Gamma_{\reflectbox{$\mathsf{Z}$}}^q$ can be obtained from the Z-channel $\Gamma_Z^q$ by a bijective mapping. The labelling of the nodes at the input and the output is simply in reversed order. \qed

\begin{figure}
	\centering
	\begin{tikzpicture}
	\draw (0,0) node [left] {$0$} -- (2,0) node [right] {$0$};
	\draw (0,-1) -- (2,0);
	\draw (0,-1) node [left] {$1$} -- (2,-1) node [right] {$1$};
	\draw (0,-2) -- (2,-1);
	\draw (0,-2) node [left] {$2$} -- (2,-2) node [right] {$2$};
	\draw (0,-3.5) -- (0.75, -3.125);
	\node at (1, -2.75) {$\vdots$};
	\node at (-0.25, -2.75) {$\vdots$};
	\node at (2.25,-2.75) {$\vdots$};
	\draw (0,-3.5) node [left] {$q-2$} -- (2,-3.5) node [right] {$q-2$};
	\draw (0,-4.5) -- (2,-3.5);
	\draw (0,-4.5) node [left] {$q-1$} -- (2,-4.5) node [right] {$q-1$};
	
	\draw (4.5,0) node [left] {$0$} -- (6.5,0) node [right] {$0$};
	\draw (4.5,0) -- (6.5,-1);
	\draw (4.5,-1) node [left] {$1$} -- (6.5,-1) node [right] {$1$};
	\draw (4.5,-1) -- (6.5,-2);
	\draw (4.5,-2) node [left] {$2$} -- (6.5,-2) node [right] {$2$};
	\draw (4.5,-2) -- (5.25, -2.375);
	\node at (4.25, -2.75) {$\vdots$};
	\node at (5.5, -2.75) {$\vdots$};
	\node at (6.75,-2.75) {$\vdots$};
	\draw (4.5,-3.5) node [left] {$q-2$} -- (6.5,-3.5) node [right] {$q-2$};
	\draw (4.5,-3.5) -- (6.5,-4.5);
	\draw (4.5,-4.5) node [left] {$q-1$} -- (6.5,-4.5) node [right] {$q-1$};
	\end{tikzpicture}
	\caption{Generalized Z-channel and generalized inverse Z-channel}
	\label{fig:Z_channel}
\end{figure}
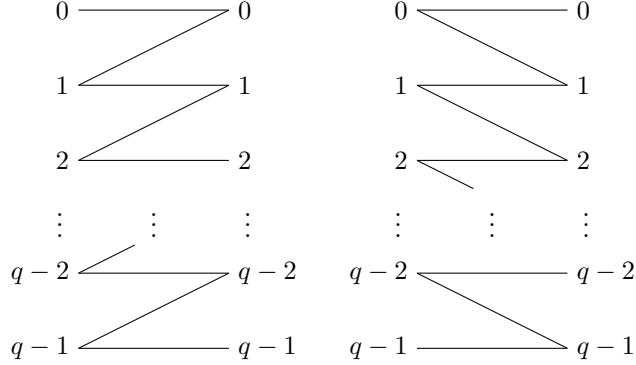

Both channels are depicted in Figure \ref{fig:Z_channel}. Combined they form a unidirectional channel which we denote as $\Gamma_{U}^q$.

The symbols $0$ and $q-1$ are of special interest for the generalized Z-channel and the inverse generalized Z-channel. For the generalized Z-channel, the symbol $0$ has the property that the sender knows that this symbol cannot be changed by the channel. If the symbol $q-1$ is received, the receiver knows that the transmitter indeed sent this symbol. The properties of $0$ and $q-1$ are swapped for the inverse generalized Z-channel compared to the generalized Z-channel. All other symbols do not have these special properties.

\section{Zero-error capacity of the generalized Z-channel}
The method we use to obtain the zero error capacity of the generalized Z-channel was already introduced by Shannon in~\cite[Theorem 7]{S56}.

The idea is to split the input symbols into sets according to their connected outputs.

\begin{Theorem}[Shannon \cite{S56}]\label{th:C_{0,q}}
	Let the set $\mathcal{S}_j$ denote the set of input symbols which are connected to the channel output $j \in \{0,\dots,q-1\}$ and let $P_i$ denote the probability of symbol $i$ at the input of the channel. Let $P_o$ be defined as
	
	\begin{equation}\label{eq:probability_capacity}
	P_o := \min_{P_i} \max_{j} \sum_{i \in \mathcal{S}_j} P_i
	\end{equation}
	over all possible input distributions $P_i$.
	
	The zero error capacity $C_{0,q}^f$ is then obtained by
	\begin{equation}
	C_{0,q}^f = \log_q\left(\frac{1}{P_o}\right) \enspace .
	\end{equation}
\end{Theorem}

Theorem \ref{th:C_{0,q}} suggests that in order to find the zero error capacity, it is possible to find the input distribution which minimizes equation \eqref{eq:probability_capacity} and to take the logarithm of the reciprocal of the result.

\begin{Property}\label{cor:zero_error_capacity}
	The zero error capacity of the generalized Z-channel and the inverse generalized Z-channel is
	\begin{equation}
	C_{0,q}^f(\Gamma_Z^q) = C_{0,q}^f(\Gamma_{\reflectbox{$\mathsf{Z}$}}^q) = \log_q\left(\left\lceil\frac{q}{2}\right\rceil\right)\,.
	\end{equation}
\end{Property}
{\bf Proof:} We apply Theorem \ref{th:C_{0,q}} to the inverse generalized Z-channel $\Gamma_{\reflectbox{$\mathsf{Z}$}}^q$ shown in Figure \ref{fig:Z_channel}.
\begin{align*}
\mathcal{S}_0 &= \{0\} \\
\mathcal{S}_1 &= \{0,1\} \\
\mathcal{S}_2 &= \{1,2\} \\
&\vdots \\
\mathcal{S}_{q-1} &= \{q-2, q-1\}
\end{align*}

Consider now an arbitrary distribution on the set of input letters \\$\{0,\dots, q-1\}$. It is obvious that
\begin{equation}
\sum_{i \in \mathcal{S}_0} P_i \leq \sum_{i \in \mathcal{S}_1} P_i
\end{equation}
with equality if and only if $P_1=0$. Assuming that $P_1 > 0$, we get an input probability which is at least as good as before according to equation \eqref{eq:probability_capacity} if we change the input probability to 
\begin{align*}
P_0' &= P_0 + P_1\\ 
P_1' &= 0\\
P_2'&= P_2\\
&\vdots \\
P_{q-1}' &= P_{q-1}\,.
\end{align*}

This is because $\sum_{i \in \mathcal{S}_2} P_i = P_1 + P_2$ can only get smaller, while for $j\in \{3,\dots, q-1\}, \sum_{i \in \mathcal{S}_j} P_i$ remains unchanged. Now we consider the distribution $P'$. As $P_1' = 0$, the set $\mathcal{S}_2$ effectively only contains the element $2$. As $\mathcal{S}_3 = \{2,3\}$ we can use the same method as before to show that the zero error capacity cannot get smaller if we change the distribution such that $P_2'' = P_2' + P_3'$ and $P_3' = 0$ while all other input probabilities remain the same.

Using this procedure inductively gives us the result that it is optimal to only have non-zero probability for every second input symbol. Looking at equation \eqref{eq:probability_capacity} it is easy to see that the optimal input distribution is the uniform distribution over all input symbols $k$ fulfilling $k \equiv 0 \mod 2$. Therefore, the zero error capacity of the inverse generalized Z-channel is

\begin{equation}
C_{0,q}^f(\Gamma_{\reflectbox{$\mathsf{Z}$}}^q) = \log_q\left(\left\lceil\frac{q}{2}\right\rceil\right) \,.
\end{equation}

Remark~\ref{rem:Z_inverse_Z} implies that the zero error capacities of the generalized Z-channel and the inverse generalized Z-channel are equal.
\qed

Furthermore, Property~\ref{cor:zero_error_capacity} can be used to show the following. 
\begin{Corollary}\label{th:zero_error_unidirectional}
    It holds $C_{0,q}^f(\Gamma_U^q) = \log_q\left(\left\lceil\frac{q}{2}\right\rceil\right)$.
\end{Corollary}
{\bf Proof:} Because of Property~\ref{cor:zero_error_capacity}, it holds that $C_{0,q}^f(\Gamma_Z^q) = C_{0,q}^f(\Gamma_{\reflectbox{$\mathsf{Z}$}}^q) = \log_q\left(\left\lceil\frac{q}{2}\right\rceil\right)$. Therefore, $\C_{0,q}^f(\Gamma_U^q) \leq \log_q\left(\left\lceil\frac{q}{2}\right\rceil\right)$. It remains to be shown that $C_{0,q}^f(\Gamma_U^q) \geq \log_q \left(\left\lceil \frac{q}{2} \right\rceil \right)$.
In order to prove this, we propose the following encoding and  strategies.

For this we define the set $\mathcal{S}:=\{ k \in \{0,\dots,q-1\} : k \equiv 0  \pmod 2 \}$. 

The encoding function
\begin{equation}
    c: \mathcal{M} \rightarrow \mathcal{S}^{n-1}
\end{equation}
can be any bijective function between $\mathcal{M}$ and $\mathcal{S}^{n-1}$.
If the sender would like to send message $m \in \mathcal{M}$, he computes $x^{n-1} = c(m)$ and sends it over the channel.
In order to send the last symbol, the sender distinguishes two cases:
\begin{enumerate}
    \item The channel caused an error within at least one of the previously transmitted symbols. In this case the sender knows the channel and sends $x_n = 0$ if the active channel is $\Gamma_Z^q$ and $q-1$ otherwise.
    \item No error has occurred within the transmission. In this case the sender sends $x_n = 0$.
\end{enumerate}

We denote the vector of symbols at the channel output as $y^n$. In order to determine which channel was active during the transmission, the decoder checks the value of $y_n$. There are three cases to be distinguished:
\begin{enumerate}
    \item $y_n = 0$: Either no transmission error occurred at all or the active channel was $\Gamma_Z^q$.
    \item $y_n = q-1$: The active channel was $\Gamma_{\reflectbox{$\mathsf{Z}$}}^q$.
    \item $y_n = 1$: No error occurred within the first $n-1$ symbols and the active channel was $\Gamma_{\reflectbox{$\mathsf{Z}$}}^q$.
\end{enumerate}
Let $\mathcal{K}_i$ denote the set of possible outputs for the channel if the symbol $i$ is transmitted. The sets $\mathcal{K}_i$ for different inputs are disjoint for $\Gamma_Z^q$ and $\Gamma_{\reflectbox{$\mathsf{Z}$}}^q$, respectively. From the last symbol $y_n$, the decoder knows which channel has been active. Therefore our coding strategy is successful for a message set $\mathcal{M}$ of size
\begin{equation*}
    |\mathcal{M}| = \left\lceil \frac{q}{2} \right\rceil ^{n-1}
\end{equation*}
achieving asymptotically the rate
\begin{equation*}
    R:=  \log_q\left(\left\lceil\frac{q}{2}\right\rceil\right).
\end{equation*}


\section{An upper bound on capacity error functions}
The analysis in the previous section deals with the zero error capacity, assuming that the number of errors in a block of length $n$ can be from $0$ to $n$. This section deals with the problem of allowing the channel to only change a fraction of the symbols during the transmission. Let this fraction be denoted as $\tau = \frac{t}{n}$, where $t$ denotes the number of possible errors.

\begin{figure}
    \centering
    \scalebox{.9}{
	\begin{tikzpicture}
	\draw [dashed] (0,1) node [left] {$*$} -- (5,1) node [right] {$*$};
	\draw [dashed] (0,0) -- (5,1);
	\draw [dashed] (5,-4.25) -- (0,1);
	\draw (0,0) node [left] {$0$} -- (5,0) node [right] {$0$};
	\draw (0,-1) -- (5,0);
	\draw (0,-1) node [left] {$1$} -- (5,-1) node [right] {$1$};
	\draw (0,-2) -- (5,-1);
	\draw (0,-2) node [left] {$2$} -- (5,-2) node [right] {$2$};
	\draw (5,-2) -- (3.75, -2.25);
	\node at (-0.25, -2.5) {$\vdots$};
	\node at (5.25,-2.5) {$\vdots$};
	\draw (0,-3.25) node [left] {$q-2$} -- (5,-3.25) node [right] {$q-2$};
	\draw (0,-4.25) -- (5,-3.25);
	\draw (0,-4.25) node [left] {$q-1$} -- (5,-4.25) node [right] {$q-1$};
	\end{tikzpicture}}
	\caption{Bipartite graph $\Gamma^{q*}$}
	\label{fig:bipartite_graph}
\end{figure}
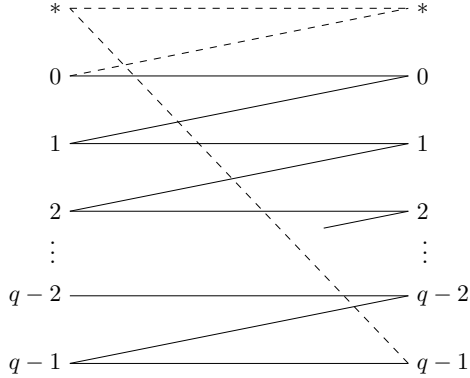

An upper bound on the capacity error function of the Z-channel can be obtained by using an upper bound on the maximal set of messages, which can successfully be transmitted over the channel $\Gamma^{q*}$ depicted in Figure \ref{fig:bipartite_graph}. 

\begin{Definition}
	The channel $\Gamma^{q*}$ is specified by the graph with nodes $\mathcal{V}_{in} = \mathcal{V}_{out} = \{*\} \cup \{0,\dots,q-1\}$ and the set of edges $\mathcal{E} = \{(*,*), (i,i) : i \in \{0,\dots,q-1 \}\} \cup \{(i,i+1): i \in \{0, \dots, q-2\}\}\cup \{(*,q-1),(0,*) \} \}$.
\end{Definition}

Let $\mathcal{M}_{q}^Z (n,t)$ denote a maximal set of messages which can successfully be transmitted over $\Gamma_Z^q$, let $\mathcal{M}_q^U(n,t)$ denote the maximal set of messages which can successfully be transmitted over $\Gamma_U^q$ and let $\mathcal{M}_{q*}(n,t)$ denote the maximal set of messages which can successfuly be transmitted over $\Gamma^{q*}$.

\begin{Theorem}\label{th:upper_bound_cap_error_function}
		\begin{equation*}\label{eq:upper_bound_q}
	C_q^f(\Gamma_U^q, \tau) \leq 1 + h_q\left(\min\left(\tau,\frac{1}{q+1}\right)\right) - \min\left(\tau,\frac{1}{q+1}\right) - h_q\left(\min\left(\tau,\frac{1}{2}\right)\right),
\end{equation*}
where $h_q$ denotes the binary entropy function with logarithms to the base $q$.
\end{Theorem}
{\bf Proof:} We will show that the following inequalities hold:
	\begin{equation}\label{eq:message_sets}
	\mathcal{M}_q^U(n,t) \stackrel{(a)}{\leq} \mathcal{M}_q^Z(n,t) \stackrel{(b)}{\leq} \mathcal{M}_q^{upper}(n,t) \enspace ,
	\end{equation}
where \begin{equation}\label{eq:M_upper}
\mathcal{M}_q^{upper}(n,t) := \frac{\sum_{i=0}^t \binom{n}{i} q^{n-i}}{\sum_{i=0}^t \binom{n}{i}} \enspace .
\end{equation}	
$(a)$ follows because every successful algorithm for $\Gamma_U^q$ is also a successful algorithm for $\Gamma_Z^q$. In order to show $(b)$, we first note that every successful strategy on $\Gamma_Z^q$ is also successful on $\Gamma^{q*}$ by not using the symbol * at the input and interpreting each * at the output as $0$. At the output of $\Gamma^{q*}$, there are at most $t$ $*$-symbols using such an encoding strategy. In order to compute the cardinality of the set of output sequences $\mathcal{S}$, they are partitioned into the sets $\mathcal{S}_0,\dots,\mathcal{S}_t$, where $\mathcal{S}_i$ denotes the set of output sequences containing $*$ exactly $i$ times. The cardinality of these sets is
\begin{equation}
    \lvert\mathcal{S}_i\rvert = \binom{n}{i} q^{n-i} \enspace .
\end{equation}
$(b)$ results from the fact that the $\mathcal{S}_i$ are disjoint and a sphere packing argument.

We define
\begin{equation*}
    C_q^{upper}(\tau) := 1 + h_q\left(\min\left(\tau,\frac{1}{q+1}\right)\right) - \min\left(\tau,\frac{1}{q+1}\right) - h_q\left(\min\left(\tau,\frac{1}{2}\right)\right) \enspace .
\end{equation*}
Next we show that $\lim_{n \to \infty} \frac{M_q^{upper}(n,t)}{n} = C_q^{upper}(\tau)$.

This limit is determined by the largest terms in both sums of equation~\eqref{eq:M_upper}.

The largest term of the sum in the denominator is equal to $\binom{n}{\min(t,n/2}$.
In order to deal with the sum in the numerator, we use Stirling's approximation to estimate the binomial coefficients of 
\begin{equation*}
    \lim_{n \to \infty} \frac{\log_q\left(\sum_{i=0}^t \binom{n}{i} q^{n-i}\right)}{n} \enspace ,
\end{equation*}
neglecting terms which are asymptotically irrelevant. The sum in the numerator omitting all asymptotically irrelevant terms and using Stirling's approximation is given by 
\begin{equation}
    q^n \sum_{i=0}^t q^{n(h_q(\xi_i)-\xi_i)} \leq q^n \tau n \max_{\xi_i \in [0,\tau]} q^{n (h_q(\xi_i) - \xi_i)} \enspace ,
\end{equation}
where we defined $i:=\xi_i n$ and $\tau:=\frac{t}{n}$.

Using elementary differential calculus, one obtains that the exponent is maximized for $\xi_i = \min\left(\tau,\frac{1}{q+1}\right)$.

Therefore,
\begin{equation}
    \lim_{n \to \infty} \frac{\log_q(\sum_{i=0}^t \binom{n}{i} q^{n-i})}{n} = 1 + h_q\left(\min\left(\tau,\frac{1}{q+1}\right)\right)-\min\left(\tau,\frac{1}{q+1}\right) \enspace .
\end{equation}

As $n$ goes to infinity, the error due to Stirling's approximation vanishes and using similar arguments for the sum in the denominator, we get
\begin{eqnarray*}
    &&\lim_{n \to \infty}\frac{\log_q (\mathcal{M}_{q}^{upper}(n,t))}{n}\\ 
    &&=
    1 + h_q\left(\min\left(\tau,\frac{1}{q+1}\right)\right) - \min\left(\tau,\frac{1}{q+1}\right) - h_q\left(\min\left(\tau,\frac{1}{2}\right)\right) \enspace ,
\end{eqnarray*}
completing the proof.
\qed
\begin{Remark}
    The proof of Theorem~\ref{th:upper_bound_cap_error_function} even shows that
    \begin{equation*}
        C_q^f(\Gamma_Z^q, \tau) \leq 1 + h_q\left(\min\left(\tau,\frac{1}{q+1}\right)\right) - \min\left(\tau,\frac{1}{q+1}\right) - h_q\left(\min\left(\tau,\frac{1}{2}\right)\right) \enspace .
    \end{equation*}
\end{Remark}

\begin{Remark}
In the theorem, we follow the idea of Hamming to get an upper bound. The new idea  
we use is to extend the channel without using all symbols at the input. Therefore we reduce
the number of output sequences, obtaining a better upper bound on the number of messages using the sphere packing argument. To the best of our knowledge, this method is new. Perhaps it can be used for other problems to get new upper bounds.
\end{Remark}



\section{Lower bound on the capacity error function}
In this section a method introduced in \cite[Lemma 2]{DL19} is recapitulated. It is then used to construct a lower bound on the capacity error function of $\Gamma_Z^q$ and $\Gamma_{\reflectbox{$\mathsf{Z}$}}^q$. In the following, the set of graphs with vertices $\mathcal{V}_{in} = \mathcal{V}_{out} = \{0,\dots,q-1\}$ all having at most degree $2$ is denoted as $\tilde{\Gamma}_2(q)$.
\begin{Lemma}[DL A\cite{DL19}] \label{lem:DLA}
	Let $R>0$ be the rate of a successful algorithm for a channel $\Delta \in \tilde{\Gamma}_2(q)$ and $\tau=1$. Then there exists a successful algorithm with rate $1-\log_q(2)$ for all $\tau$.
\end{Lemma}

\begin{Lemma}[DL B\cite{DL19}] \label{lem:DLB}
	If there exists a successful algorithm for a channel $\Delta \in \tilde{\Gamma}_2(q)$ with rate $R>0$ for all $\tau$, then there exists a successful algorithm with rate $1-h(\tau) \log_q(2)$ for $\Delta$ and $0 \leq \tau \leq \frac{1}{2}$, where $h(\cdot)$ denotes the binary entropy function.
\end{Lemma}

We omit the proof of Lemma \ref{lem:DLA} and give only the idea of the construction achieving the rate proposed in Lemma \ref{lem:DLB}. A full proof is given in \cite{DL19}.

{\bf Proof sketch:} The strategy of the proof of \cite{DL19} is to give an algorithm which can transmit $\frac{q^n}{\binom{n}{t}}$ messages. This algorithm then achieves the rate claimed in the Lemma.
The algorithm uses the first $k_1 = n - \log_q \binom{n}{t}$ symbols to transmit the information symbols and to use the rest of the block to correct potential errors caused by the channel. It is sufficient for the receiver to know the error positions because the nodes at the receiving side of the graph only have a degree of at most $2$. Let $t_1$ denote the number of errors within the information symbols and $n_1 := n - k_1 = \log_q \binom{n}{t}$.

If $\binom{k_1}{t_1} \leq q^{R n_1}$, then the successful algorithm for $\tau = 1$ can be used to transmit the error positions, and since this works for any number of errors, the receiver can decode the message correctly.
Otherwise, let $k_2 = \log_q \binom{k_1}{t_1}$ and use the interval $[k_1+1, k_1 + k_2]$ to send the error positions of the $t_1$ errors within the first interval. This process continues by defining $k_{i+1} = \log_q \binom{k_i}{t_i}$ and $n_{i+1} = n_i - k_{i+1}$ and is stopped when $\binom{k_i}{t_i} \leq q^{R n_i}$. Then the successful algorithm for $\tau = 1$ can be used to transmit the error positions with certainty. In the last block, the $k_i$s have to be transmitted as well. The amount of symbols that need to be reserved in order to do this is of order $\mathcal{O}(\log(n)\log(t))$. This does not change the asymptotically achievable rate of the algorithm.
\qed

The capacity error function for the Z-channel $\Gamma_Z^q$ with an additional edge $(0,q-1)$ was analyzed by Deppe and Lebedev in \cite{DL19} for $q>3$. We denote this channel as $\Gamma^q_{DL}$ and its error capacity function as $C_q^f(\Gamma^q_{DL},\tau)$.
Since the additional edge can only make the channel worse,

\begin{equation}\label{eq:upper_bound_wrap}
C_{q}^f(\Gamma_Z^q,\tau) \geq C_q^f(\Gamma^q_{DL},\tau).
\end{equation}

We denote the asymptotic rate of the algorithm achieving $C_q^f(\Gamma_{DL}^q,\tau)$ as
\begin{equation}\label{eq:asymp_DL}
    R_{DL}^q:= \begin{cases}
1- h(\tau )\log_q 2 & 
\text{if } 0\leq \tau \leq \frac{1}2\\
1-\log_q 2 & \text{if } \tau > \frac 12.
\end{cases}
\end{equation}

The algorithm achieving $R_{DL}^q$ uses the methods of the proofs of Lemma~\ref{lem:DLA} and Lemma~\ref{lem:DLB}.
The algorithm in the proof of Lemma~\ref{lem:DLB} requires the possibility of error-free transmission at a positive rate for $\tau=1$. For the channel $\Gamma_{DL}^3$, this is not possible. However for $\Gamma_U^3$ the proposed method works, as it is possible to transmit data with a positive rate for $\tau=1$ by Corollary~\ref{th:zero_error_unidirectional}.

\begin{Corollary}\label{cor:lower_bound}
	Let $q \geq 3$. Then the capacity error functions of $\Gamma_Z^{q}$, $\Gamma_{\reflectbox{$\mathsf{Z}$}}^{q}$ and $\Gamma_U^{q}$ are lower bounded by $R_{DL}^q$.
	
	\begin{equation}
	C_{q}^f(\Gamma_Z^{q},\tau) \geq C_{q}^f(\Gamma_{U}^{q},\tau) \geq R_{DL}^q
	\end{equation}
\end{Corollary}

We note that Corollary \ref{cor:lower_bound} does not give information on the capacity error functions for the binary channels.

\begin{Corollary}\label{cor:odd_q}
    For odd $q$ the following equations hold for $\tau \geq \frac{1}{2}$:
    \begin{equation}
        C_q^f(\Gamma_Z^q,\tau) = C_q^f(\Gamma_U^q,\tau) = C_{0,q}^f(\Gamma_Z^q) = C_{0,q}^f(\Gamma_U^q)
    \end{equation}
\end{Corollary}
{\bf Proof:} In accordance with the definition in the proof of Theorem~\ref{th:upper_bound_cap_error_function}, it holds that $C_q^{upper}(\Gamma_Z^q,\tau) = \log_q\left(\frac{q+1}{2}\right)$ for $\tau \geq \frac{1}{2}$, which is equal to $\log_q(\lceil \frac{q}{2} \rceil) = C_{0,q}^f(\Gamma_U^q)$ for odd $q$ by Property~\ref{cor:zero_error_capacity}.
\qed

\begin{figure}
    \centering
    \scalebox{0.65}{
    \input{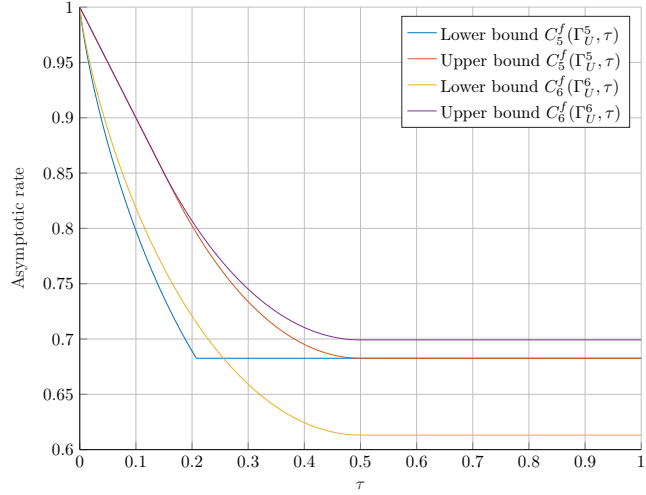}}
    \caption{Lower and upper bound on the capacity error function of the unidirectional channels $\Gamma_U^5$ and $\Gamma_U^6$}
    \label{fig:bounds_capacity_error_function}
\end{figure}

Figure \ref{fig:bounds_capacity_error_function} illustrates the results obtained in Theorem~\ref{th:upper_bound_cap_error_function}, Property~\ref{cor:zero_error_capacity} and Corollaries ~\ref{cor:lower_bound} and~\ref{cor:odd_q}. Property \ref{cor:zero_error_capacity} shows that the methodology proposed in \cite{DL19} (Lemma~\ref{lem:DLA} in this work) only achieves the zero error capacity of $\Gamma_Z^q$ if $q$ is even. However, the strategy proposed in the proof of Corollary~\ref{th:zero_error_unidirectional} is able to achieve the zero error capacity for $\Gamma_U^q$ and $\Gamma_Z^q$ for any $q$. The results in Figure~\ref{fig:bounds_capacity_error_function} are in accordance with Corollary~\ref{cor:odd_q} for $q=5$. For even $q$ we only know the zero error capacity which is achieved by the lower bound. For $\tau \neq 1$ our results do not show whether upper and lower bounds on $C_q^f(\Gamma_U^q,\tau)$ are tight. $q=6$ was chosen as an example for even $q$.


\section{Constructing lower bounds using rubber methods}
The rubber method was developed in \cite{ADL05}. It is used to determine a lower bound on the capacity error function of a channel. Basically it reserves a sequence of symbols to notify the receiver that his previously received symbol has been an erroneous one. Therefore, after receiving the rubber sequence, the receiver deletes the rubber plus the previous symbol. 
In \cite{ADL05}, a rubber sequence consists of $r$ equal rubber symbols $b\in\Q$.
The sender then retransmits the previously erroneous symbol again. Because the number of errors $t$ is fixed and smaller than the blocklength $n$, he will succeed after at most $t$ trials. In \cite{ADL05} it is shown that that the rubber method is a successful algorithm.
This means the decoder is able to decode each message correctly. In the error analysis in \cite{ADL05}, the authors consider two kinds of errors:
a {\bf standard error} (which means a symbol is changed to another symbol and
the sender sends the rubber sequence) and a {\bf towards rubber error} (which means a symbol is changed to a rubber symbol such that the receiver obtains a rubber sequence). If a {\bf towards rubber error} occurs, a correctly received symbol is deleted and has to be sent again.
For the generalized Z-channel $\Gamma_Z^q$, a {\bf towards rubber error}
is not possible if we use $r$ times $q-1$ as the rubber sequence. Also, for the
inverse generalized Z-channel $\Gamma_{\reflectbox{$\mathsf{Z}$}}^q$, a {\bf towards rubber error} is not possible if we use $r$ times $0$ as the rubber sequence. Therefore, the sender does not have to retransmit the previously erroneous symbol again, because for those channels there is only the possibility of a {\bf standard error}. We denote this modified algorithm (without retransmissions) by $A(r,b)$ if we use
$b^r=(b,\dots, b)$ as the rubber and we get:
\begin{Lemma}\label{lem:rubber_retransmission}
	The modified rubber strategy $A(r,q-1)$ [$A(r,0)$] is a successful algorithm
	for the the generalized [inverse] Z-channel $\Gamma_Z^q$ [$\Gamma_{\reflectbox{$\mathsf{Z}$}}^q$].
\end{Lemma}

In the following, we denote this rubber method without retransmission as the modified rubber method.

To calculate the rate of this algorithm we need, as in \cite{ADL05}, the following definitions. Let $z_r^{r+1}=qz^r_r-q+ 1$. It is well known that for  $n \rightarrow \infty $ the number of sequences of length $n$ not containing a block $b^r=(b,\dots, b)$ is asymptotically equal to $z_r^{n}$ (see
\cite{ADL05, B77}, how to choose the initial value for the iteration).

\begin{Theorem}\label{th:rubber_zr}
	Let $z_r$ be defined as above. Then for the generalized [inverse] Z-channel $\Gamma_Z^q$ [$\Gamma_{\reflectbox{$\mathsf{Z}$}}^q$] for $q \geq 2$ we get 
	\begin{equation*}
	C_q^f(\Gamma_Z^q,\tau) \geq R_{mr}:= \begin{cases}
	\max\limits_{2\leq r\in\NN} (1-r\tau) \log_q z_r& \text{if } 0 \leq \tau \leq \frac{1}{2}\\
	0 & \text{if } \tau > \frac{1}{2}\, .
	\end{cases}
	\end{equation*}
\end{Theorem}
{\bf Proof:} The modified $r$-rubber method can be used to transmit information vectors not containing the rubber sequence. The blocklength is $n$ and the number of erroneous symbols is at most $t$. For each error we require $r$ symbols for correction. Therefore the length of the information vectors can be at most $n-rt$. The number of these sequences is asymptotically equal to $z_r^{n-rt}$. Therefore the asymptotic rate achieved by the modified $r$-rubber method is
\begin{equation*}
(1-r\tau) \log_q z_r.
\end{equation*}
This gives the following lower bound on the capacity error function
\begin{equation*}
C_q^f(\Gamma_Z^q,\tau ) \geq (1-r\tau) \log_q z_r. 
\end{equation*}
We use the modified rubber method with the rubber length $r$ that achieves the highest asymptotic for the respective value of $\tau$ and obtain the lower bound $R_{mr}$. \qed

\begin{Remark}\label{rem:rubber}
	For the generalized Z-channel $\Gamma_Z^q$, the modified rubber method with a single $0$ as the rubber symbol achieves the rate
	\begin{equation}
	R = (1-\tau) \log_q(q-1)\,.
	\end{equation}
\end{Remark}
{\bf Proof:} Using Lemma~\ref{lem:rubber_retransmission}, it is easy to see that for every erroneous symbol one extra symbol has to be transmitted in order to achieve error free transmission. Therefore it is possible to achieve a message set $\mathcal{M}$ of cardinality

\begin{equation}
M = (q-1)^{n-t}\, ,
\end{equation}
leading to a rate
\begin{equation}
R = (1-\tau) \log_q(q-1)\,.
\end{equation}\qed

%
By an adjustment of the modified rubber method for the channels $\Gamma_Z^q$ and $\Gamma_{\reflectbox{$\mathsf{Z}$}}^q$ we obtain the following result:

\begin{Theorem}\label{th:rubber_unidirectional}
	Let $\Gamma_U^q$ be a unidirectional channel consisting of the generalized Z-channel $\Gamma_Z^q$ and the inverse generalized Z-channel $\Gamma_{\reflectbox{$\mathsf{Z}$}}^q$ ($q \geq 2$). Then
	we have 
	\[
	C_q^f(\Gamma_U^q,\tau)\geq R_{mr}.
	\]
\end{Theorem}
{\bf Proof:} We adapt the modified rubber method to make it usable for $\Gamma_U^q$. Furthermore, we show that this method is successful and achieves the same asymptotic rate as the modified rubber method. The sender starts by using the encoding strategy for the generalized Z-channel $\Gamma_Z^q$ and then adapts if he discovers that the active channel is $\Gamma_{\reflectbox{$\mathsf{Z}$}}^q$. After all the data symbols are transmitted, an extra symbol (either $0$ or $q-1$ according to the channel) is added to tell the receiver which channel was present. The receiver can thus adjust his decoding algorithm accordingly. The previously described steps are now elaborated in more detail.

The only additional difficulty compared to the situation in Theorem~\ref{lem:rubber_retransmission} is that it is not known to the sender and the receiver whether $\Gamma_Z^q$ or $\Gamma_{\reflectbox{$\mathsf{Z}$}}^q$ is used for the transmission of each codeword but once the first error occurs the sender knows which channel is in action.

At the start of the transmission, the sender assumes that $\Gamma_Z^q$ is the active channel of $\Gamma_U^q$. He therefore uses the modified rubber method for $\Gamma_Z^q$ of length $r$ as his encoding strategy.

When the channel inflicts the first error to the transmission, the sender knows which channel is active. We denote the transmitted symbol corresponding to the first error with $x_{fe}$. According to the actual channel, the sender adjusts his encoding strategy henceforth.
We distinguish two cases:
\begin{enumerate}
	\item The active channel is $\Gamma_Z^q$. In this case the channel keeps its encoding strategy as it is. After the sender knows that the all information symbols can be correctly retrieved by the decoder, he fills the remaining block with $0$ symbols. Notice that this last step involves knowledge about the decoding algorithm at the sender.
	\item The active channel is $\Gamma_{\reflectbox{$\mathsf{Z}$}}^q$. In this case there are two cases to be distinguished
	\begin{enumerate}
	    \item A {\bf standard error} occurred: In this case the transmitter sends $r$ times the symbol $q-1$. If there are not enough symbols left in the block, the sender sends $q-1$ for the remaining symbols.
	    \item a {\bf towards rubber error} occurred: In this case the transmitter sends $r-1$ times the symbol $q-1$. If there are not enough symbols left in the block, the sender sends $q-1$ for the remaining symbols.
	\end{enumerate}
	The modified rubber method for $\Gamma_{\reflectbox{$\mathsf{Z}$}}^q$ is used from now on to retransmit $x_{fe}$ and to send the remaining information symbols. But before sending them, they are adapted to the channel $\Gamma_{\reflectbox{$\mathsf{Z}$}}^q$ by using the mapping
	\begin{align*}
        f&: \mathcal{Q} \to \mathcal{Q}\\
        k &\mapsto k+1 \mod q.
    \end{align*}
    Potential remaining symbols within the block are filled by sending $q-1$ symbols.
\end{enumerate}

Let the received sequence be denoted as $y^n=(y_1,\dots,y_n)$. The decoding strategy depends on the symbol $y_n$:
\begin{enumerate}
    \item $y_n=0$: The decoder uses the decoding procedure for the modified rubber method for $\Gamma_Z^q$ until he retrieves the message $m$. Potential remaining symbols within the block are ignored by the decoder.
    \item $y_n=q-1$: The decoder checks for the first occurrence of the sequence $r$ times $q-1$. All previously sent symbols, with exception of the symbol that was sent right before this sequence, have been received correctly. If those information symbols are sufficient to obtain the message with certainty, the decoder outputs the message $m$. Otherwise, the receiver continues its decoding procedure by using the decoding algorithm of the modified rubber method for $\Gamma_{\reflectbox{$\mathsf{Z}$}}^q$ on the remaining symbols. The function
    \begin{align*}
        f&^{-1}: \mathcal{Q} \to \mathcal{Q}\\
        k &\mapsto k-1 \mod q
    \end{align*}
    is then to be applied on all information symbols which were sent after the first occurrence of the sequence $r$ times $q-1$ to obtain the information vector. This vector uniquely determines the message $m$.
\end{enumerate}
If the first error occurs within the last $r$ symbols, this error cannot be corrected anymore, but the receiver knows that all previous symbols have been received correctly. Those contain the information about the message, and thus this case is not problematic for successful decoding.
\qed

The following result follows directly from Remark~\ref{rem:rubber} and Theorem~\ref{th:rubber_unidirectional}.

\begin{Corollary}
    The asymptotic rate achieved by using the modified rubber method on the unidirectional channel $\Gamma_U^q$ using a single symbol rubber is
	\begin{equation}
	R = (1-\tau)\log_q (q-1)\,.
	\end{equation}
\end{Corollary}

\begin{Remark}
	The values for $z_r$ in Theorem~\ref{th:rubber_zr} and Theorem~\ref{th:rubber_unidirectional} 
	can be computed. An important case is $q=2$.
	It was proved that 
	\[
	C_2^f(\Gamma^2,\tau)=(1-3\tau)\log_2 \left(\frac{1+\sqrt{5}}{2} \right)\] 
	for \((3+\sqrt{5})^{-1} \leq \tau \leq 1/3\) 
	(see equation~\eqref{eq:symmetric_channel}). 
	In \cite{ADL05} it was shown that this capacity can be 
	achieved with the rubber method.
\end{Remark}

\begin{figure}
	\centering
	\scalebox{0.65}{
		\input{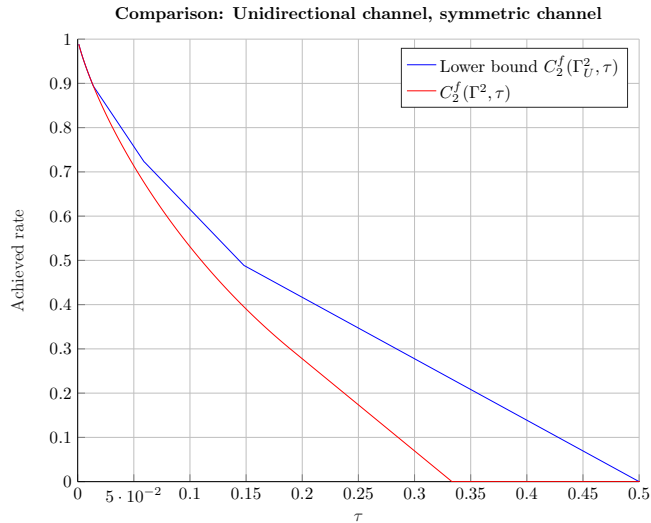}}
	\caption{Comparison of the lower bound on $C_2^f(\Gamma_U^2,\tau)$ obtained by using the modified rubber method and the capacity error function of the symmetric channel $C_2^f(\Gamma^2,\tau)$}
	\label{fig:comparison_Berlekamp}
	\vspace{-1em}
\end{figure}

If we apply the modified rubber method to the binary case we get the following result.
The capacity error function of the unidirectional channel is lower bounded by
\begin{equation*}
C_2^f(\Gamma_U^2,\tau) \geq \begin{cases}
\max\limits_{2\leq r\in\NN} (1-r\tau) \log_2 z_r& \text{if } 0 \leq \tau \leq \frac{1}{2}\\
0 & \text{if } \tau > \frac{1}{2}\, .
\end{cases}
\end{equation*}
The result for $r=2$ is
\[
(1-2\tau)\log_2 \left(\frac{1+\sqrt{5}}{2} \right).
\]
The major change compared to the result given in equation~\eqref{eq:symmetric_channel} is the change of the factor $(1-3\tau)$ to $(1-2\tau)$. It occurs because retransmissions after erroneous symbols are unnecessary. 
Figure~\ref{fig:comparison_Berlekamp} shows the lower bound on $C_2^f(\Gamma_U^2,\tau)$ obtained by using the modified rubber method and $C_2^f(\Gamma^2,\tau)$ for comparison. This result is different without feedback, since the capacity error functions of symmetrical and unidirectional channels are the same in that case.

\section{Conclusion}
In this work we have analyzed channels with feedback with a fixed maximal fraction of erroneous symbols. A major focus of this work has been asymmetric and unidirectional channels. Notably, we have shown that the capacity error function with feedback for the unidirectional channel composed of Z-channel and inverse Z-channel is larger than the capacity error function of the symmetric channel for a significant proportion of the possible values of $\tau$. This is not the case for the same channels without feedback.
The channels analyzed in this work have at most two input symbols connected to each output. As the knowledge of an erroneous position at the receiver implies knowing the value of the respective symbol, retransmissions are not necessary if the encoding strategy is chosen in a way such that the receiver is able to obtain the error positions. This can be used to create encoding strategies achieving a higher rate for many channels using the modified rubber method. Furthermore it was shown how to change the modified rubber method for unidirectional channels. The method proposed shows that it is possible to achieve the same rate for the unidirectional channel consisting of the generalized Z-channel and the inverse generalized Z-channel as for its components. Obtaining tighter bounds on the capacity error functions for several of the channels analyzed in this paper is an interesting topic for further research.

\section*{Acknowledgement}
The authors would like to thank Frans Willems for making them aware that the methods proposed in~\cite{DL19} might be adapted for use on unidirectional channels.

Christian Deppe was supported by the Bundesministerium 
			f\"ur Bildung und Forschung (BMBF) through Grant 16KIS1005. Vladimir Lebedev was supported in part by the Russian Foundation for Basic Research, project no. 19-01-00364. Georg Maringer’s work was supported by the German Research Foundation (Deutsche Forschungsgemeinschaft, DFG) under Grant No.WA3907/4-1.

\end{document}